\documentclass[]{aa}

\usepackage{graphicx}% Include figure files

\usepackage{bm}% bold math
\usepackage{natbib}
\usepackage{url}
\usepackage{color}
\newcommand{\CaII }{Ca\,{\footnotesize II~}}
\begin{document}
\bibliographystyle{aa}
\input epsf

\title{The solar chromosphere at high resolution with IBIS}
\subtitle{ III. Comparison of Ca II K and Ca II 854.2 nm imaging}

\author{K. P. Reardon\inst{1,2}
\and H. Uitenbroek\inst{2}
\and G. Cauzzi\inst{1,2}}

\institute{INAF - Osservatorio Astrofisico di Arcetri, 50125 Firenze, Italy
\and National Solar Observatory, P.O. Box 62, Sunspot NM 88349, USA
\mail{kreardon@arcetri.astro.it}
}
\date{\today}

\abstract
{Filtergrams obtained in \CaII H, \CaII K and H$\alpha$ are often employed as 
diagnostics of the solar chromosphere. The vastly disparate appearance 
between the typical filtergrams in these different lines calls into question the 
nature of what is actually being observed.}
{We investigate the lack of obvious structures of magnetic origin such as fibrils 
and mottles in on-disk \CaII H and K images.}
{We directly compare a temporal sequence of classical \CaII K filtergrams with a 
co-spatial and co-temporal sequence of spectrally resolved \CaII 854.2 images obtained 
with the Interferometric Bidimensional Spectrometer (IBIS), considering the 
effect of both the spectral and spatial smearing. We analyze the temporal 
behavior of the two series by means of Fourier analysis.}
{The lack of fine magnetic structuring in \CaII K filtergrams, 
even with the narrowest available filters, is due to observational 
effects, primarily contributions from the bright, photospheric 
wings of the line that swamp the small and dark chromospheric structures.
Signatures of fibrils remain however in the temporal evolution of 
the filtergrams, in particular with the evidence of magnetic shadows 
around the network elements. The \CaII K filtergrams do not appear, however, 
to properly reflect the high-frequency behavior of the chromosphere. Using the same analysis,
we find no significant chromospheric signature in the  {\it Hinode}/SOT \CaII H quiet-Sun filtergrams.}
{The picture provided by H$\alpha$ and \CaII 854.2, which show significant portions of the 
chromosphere dominated by magnetic structuring, appears to reflect the true and essential nature 
of the solar chromosphere. Data which do not resolve, spatially or spectrally, this aspect may 
misrepresent the behavior the chromosphere.}	

\maketitle
\titlerunning{Comparison of \CaII K and 854.2 nm}
\authorrunning{Reardon et al.}

\section{Introduction}

The \CaII H and K resonance spectral lines have long been used for
diagnostics of the solar chromosphere. The crucial results obtained
with these strong lines, particularly those based on observations
using slit-based spectrographs, have been 
summarized by \citet{1991SoPh..134...15R}.
%Rutten \& Uitenbroek 1991.
Imaging observations, most often using Lyot-type filters (0.03--0.2 nm FWHM), 
have also been widely employed.
Their use has much increased in recent years in part due to the availability of
efficient image stabilizing systems %(tip-tilt correctors, adaptive optics systems) 
and image reconstructions techniques,  %(speckle, phase diversity, blind deconvolution), 
both of which allow high spatial resolution to be maintained over long period of times.

Several studies have used filter observations
in an effort to assess the characteristics of the chromosphere
via the examination of the spatial distribution and temporal evolution of structures in the
two-dimensional field of view 
\citep[e.g.][]{1998ASPC..140..223E,2005A&A...441.1183D,2006A&A...459L...9W,2007A&A...462..303T}.
%Ermolli, 1998, De Wijn et al 2005, Woeger et al 2006; Tritschler et al. 2007
Most recently, a broadband \CaII H filter (0.22 nm FWHM) has been
flown on the {\it Hinode} satellite, allowing the acquisition of
long time series free from terrestrial atmospheric distortions. Novel
results from this instrument include small scale penumbral jets
\citep{2007Sci...318.1594K},
%Katsukawa Science 2007
the spatial patterning of chromospheric umbral oscillations \citep{2007PASJ...59S.631N},
%Nagashima PASJ 2007
the rapid evolution of spicules \citep{2007Sci...318.1574D}, and more.
%De Pontieu et al 2007 Science

However, the on-disk images obtained in the H and K lines
with these filters are always 
significantly different from the appearance of
the chromosphere in the other preeminent
chromospheric line, H$\alpha$. 
Observations in the latter
line, using even relatively broad filters (0.1 nm FWHM) show a highly structured 
environment including fibrils, mottles, and filaments. This is consistent with the increasing dominance 
of the magnetic field as the plasma $\beta$ drops at increasing heights in the atmosphere. 
Images in the \CaII H and K lines,
instead, do not show many examples of such structuring by the magnetic field, 
%GC  \citep{2006ASPC..354..276R,2007ASPC..368...27R}, 
even when using the narrowest available passbands \citep[cf. the images shown in][]{2006A&A...459L...9W}.
% Woeger et al. 2006
Observations of a correspondence between fibrils seen in H$\alpha$ and \CaII H and K were observed
as long ago as \citet{Bumba:1965p2863} and \citet{Beckers.1964PhDT} using spectroheliograms, 
but with a reduced contrast.
\citet{2007ASPC..368...27R} 
% Rutten 2007
remarks that the fibrils seen in \CaII H and K filtergrams are at best visible only as ghostly impressions 
of their counterparts observed in H$\alpha$.
Why does the observed scene differs so greatly between these two
sets of ``chromospheric'' lines?

Another set of \CaII  lines, the so-called ``infrared triplet'' around 850 nm may 
provide for the resolution of this apparent discrepancy.
Like H$\alpha$, the \CaII IR triplet lines are subordinate, but whereas the lower level of
the former is coupled to the hydrogen ground level via the extremely strong Ly$\alpha$
radiative transition, the lower level of the latter lines is metastable and
only coupled to the \CaII\ ground level via electronic collisions.
This accounts for a stronger coupling of the \CaII IR triplet to local conditions and
allows a more straightforward interpretation of the line formation \citep{1989A&A...213..360U}.
Observationally,
the use of the infrared lines also provides several significant
advantages with respect to
H and K, including a typically better response of digital detectors in the red, a reduction in
atmospheric turbulence, and a higher photon flux.  
Indeed, in recent years, the use of the infrared triplet for solar studies has increased notably 
\citep[see, e.g.][]{2006SoPh..235...55S,2006ApJ...639..516U,2006A&A...456..689T},
%2006ApJ...645..776U, 2007ApJ...663.1386P}, 
although studies combining spatial and spectral high resolution over extended 
fields of view (FOV) are still scarce. 

In Paper I and II of this series \citep[][]{2008A&A...480..515C,2008arXiv0807.4966V} 
%Cauzzi et al. 2008
we presented novel observations of the \CaII\ 854.2 nm line of the infrared triplet, 
obtained with the Interferometric Bidimensional Spectrometer \citep[IBIS,][]{2006SoPh..236..415C}.
%Cavallini 2006
These studies established the suitability of imaging spectroscopy in this line for high--resolution studies 
of chromospheric diagnostics. A central finding of Paper I was the nearly ubiquitous 
occurrence of fibrilar structures. These originate from even the smallest magnetic elements, and 
appear to fill large portions of the chromospheric volume, even in ``quiet'' areas. 
Their presence indicates that even at the relatively low chromospheric heights sampled by the \CaII 854.2 line, 
the atmosphere is already highly structured by the pervasive magnetic fields,
entirely consistent with the 
picture provided by H$\alpha$ images. 
The fibrils in turn play a crucial role in 
shaping the chromospheric dynamics, in particular producing a strong reduction in the oscillatory power at periods 
around 3 minutes, correlated with the absence of chromospheric acoustic shocks 
\citep[][Paper I, Paper II]{2007A&A...461L...1V}.

The question is then why this subordinate 
line would display a more ``chromospheric''  scene
than the resonance H and K
lines, which have a far larger opacity. In Paper I we put forward the
hypothesis that a large part of the problem lies in the strong
observational limitations that still affect \CaII H and K
imaging. Namely, while the chromospheric contribution of the \CaII H
and K is limited to a narrow core of less than 0.02 nm width, virtually
all the imaging is performed with much broader filters, having FWHM
passbands in the range of 0.03\,--\,0.3 nm. 
Further, 
the decreased photon flux
and low filter transmission lead to long exposures (often one second
or more) that, combined with the increased seeing disturbances at
shorter wavelengths, severely limits the spatial resolution that can
be consistently achieved. 

In this paper, we explore these and other effects by a direct comparison
between spatially and spectrally resolved IBIS \CaII 854.2 data and
cotemporal \CaII K filtergrams. This builds on the preliminary results
that have been reported in \citet{2007ASPC..368..151R} and allows us
to compare the clear chromospheric signatures seen in the IBIS spectra
with the behavior observed in the \CaII K filtergrams.

%******************************************************************************
\section{Observations} \label{s_obs}
%******************************************************************************

%******************************************************************************
\begin{figure}
\includegraphics[width=8.4cm,height=6cm]{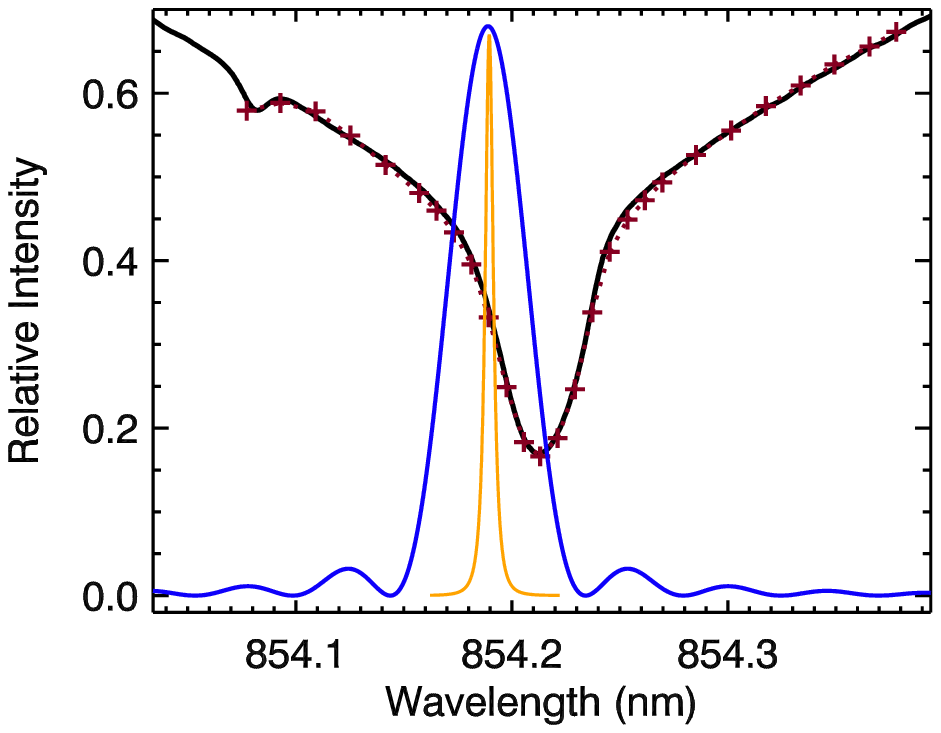}
\includegraphics[width=8.4cm,height=6cm]{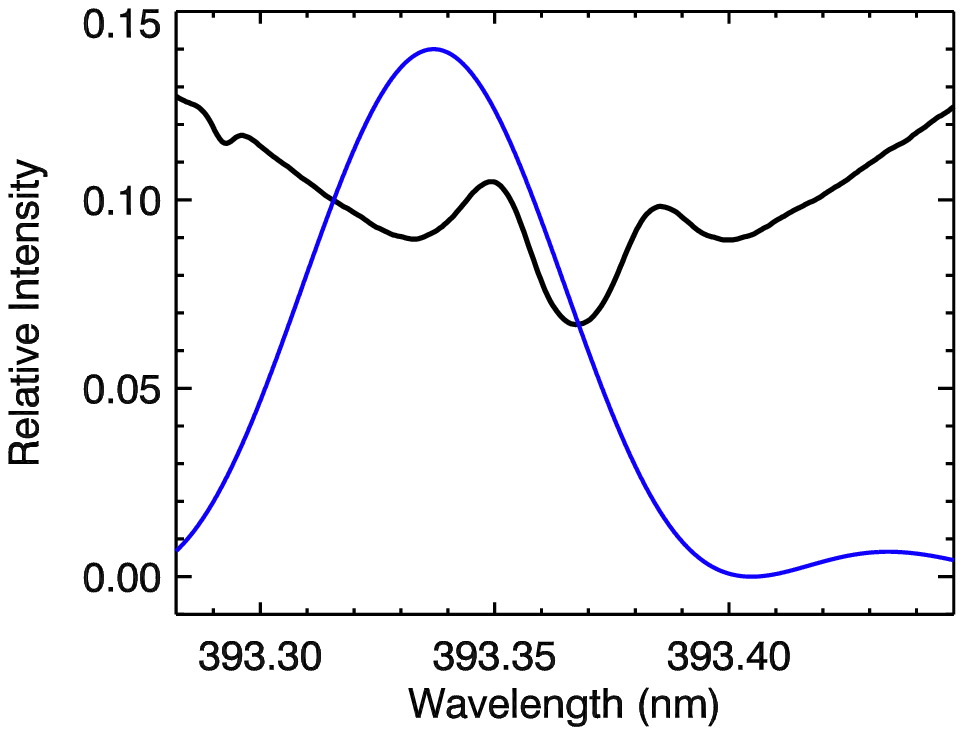}
\caption{
{\it Top panel:} \CaII 854.2 nm atlas profile ({\it black line}), the
IBIS transmission profile ({\it orange line}), and the ``synthetic''
instrumental profile ({\it blue line}), derived as described in the
text. The red line is the observed spectral profile averaged over a
portion of the field of view with the crosses indicating the nominal
spectral sampling.  {\it Bottom panel:} Atlas profile for the core of
\CaII K ({\it solid black line}) and theoretical instrumental passband
of the Halle filter during the observations ({\it blue line}).  The
plotted wavelength range is scaled according to the wavelengths of the
two lines.  In both plots, the transmission of the filter passband has
been arbitrarily scaled to fit it within the plotted range of
intensities for the atlas profile.}
\label{f_sync}
\end{figure} 
%******************************************************************************

%******************************************************************************
\begin{figure*}
\includegraphics[width=6cm,height=6cm]{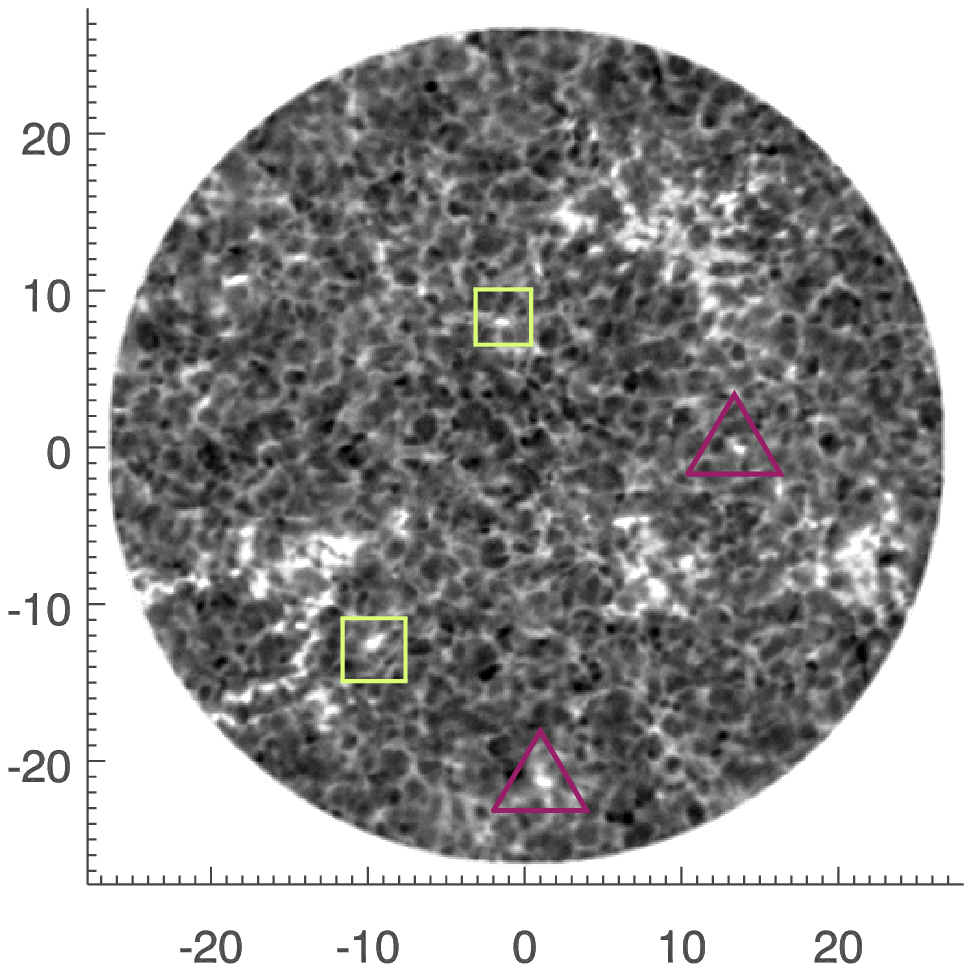}
\includegraphics[width=6cm,height=6cm]{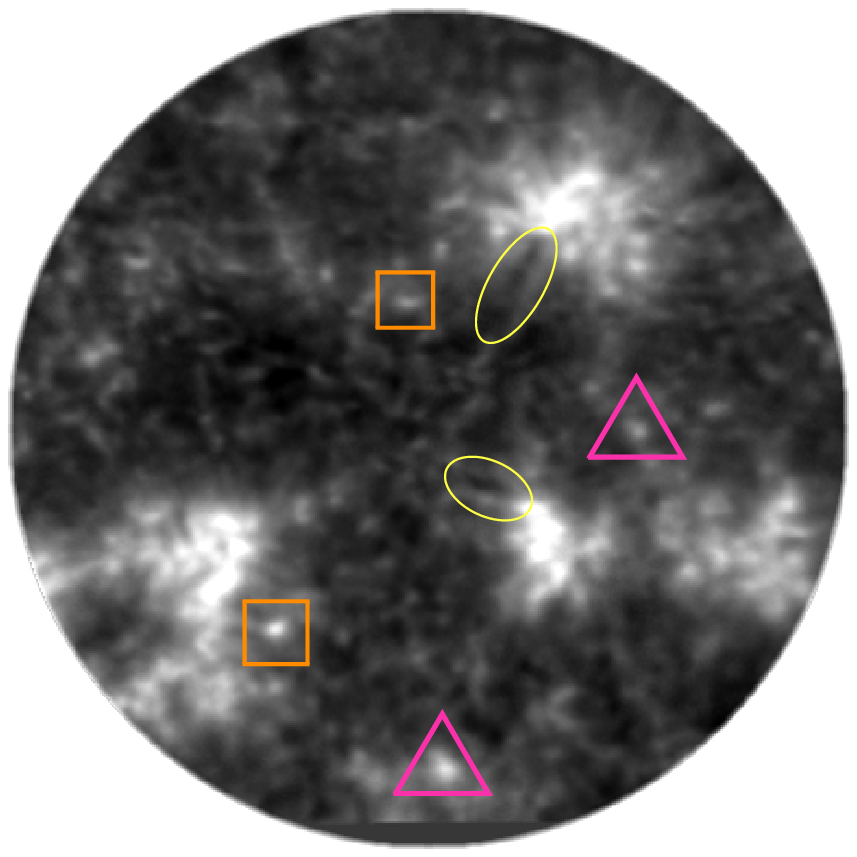}
\includegraphics[width=6cm,height=6cm]{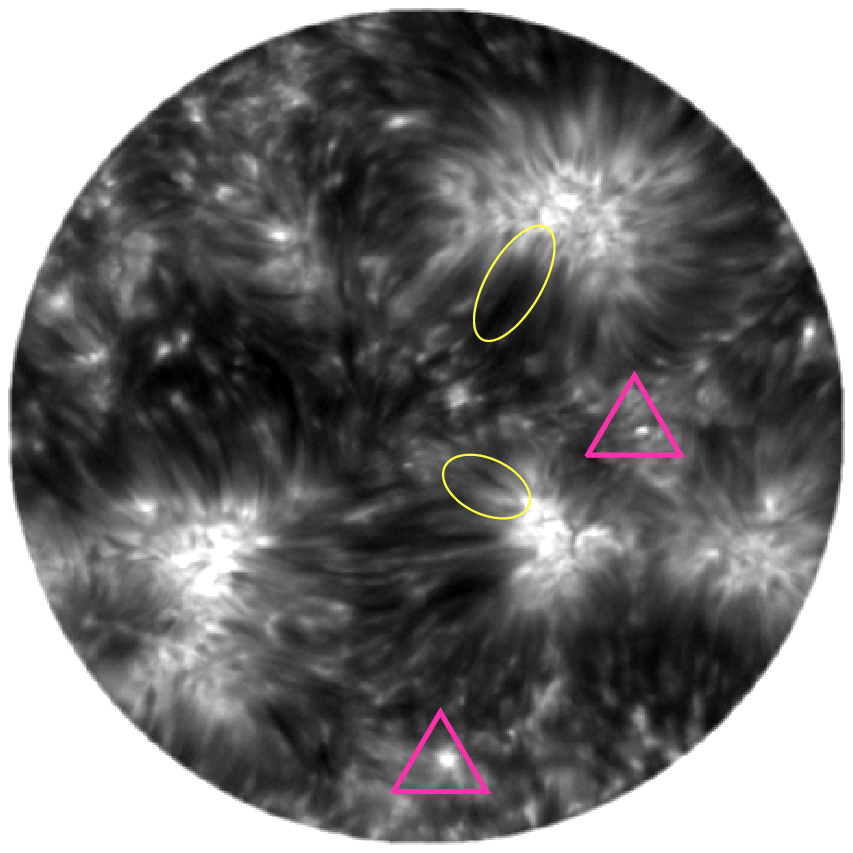}
\caption{Simultaneous individual images from the full observing sequence:
({\it left}) \CaII 854.2 nm blue wing $-0.07$ nm from line core; 
({\it center}) \CaII K filtergram; 
({\it right}) \CaII 854.2 line core.
The orange squares, yellow ovals, and pink triangles indicate, respectively, 
structures present in the \CaII line wing images, in the  \CaII line 
core images, or in both. The spatial scale is given in Mm.}
\label{f_fov}
\end{figure*} 
%******************************************************************************

The observations were obtained on 2 June 2004, starting around 15:00
UT, at the Dunn Solar Telescope (DST) of the US National Solar
Observatory.  Seeing conditions were excellent during most of the
approximately one-hour-long sequence, aided by a high order adaptive
optics system \citep{2004SPIE.5490...34R}. The target region was a
quiet area at disk center, that included some portions of enhanced,
bipolar network, well observed in the MDI high resolution magnetic
maps. At continuum wavelengths the magnetic concentrations
corresponded to a few transient pores, that appeared and disappeared
repeatedly during the course of the observations. In coronal images,
the region appeared as a fading coronal bright point.

%\subsection{Acquired Data}\label{ss_obs_ibis}

The \CaII 854.2 data utilized in this paper have been described in
detail in Paper I. We thus summarize here only the most important
aspects of the observations. The field of view (FOV) covered an 80\arcsec
diameter circle on the Sun with a plate scale of 0.166\arcsec/pixel. The
\CaII 854.2 nm line was sampled with 27 spectral points (see
Fig.~\ref{f_sync}), which were acquired in a total of 7 seconds. The
repetition time between scans was 19 seconds (the entire sequence also included scans
of two other photospheric lines). A total of 175 scans
of the spectral line were obtained for a total duration of 55 minutes.
The instrumental spectral profile of IBIS is described in \citet{2008A&A...481..897R}
%Reardon \& Cavallini 2008.
and at this wavelength has a FWHM of 4.4 pm, with spectrally scattered light
less than 1.5\% of the total transmission.

Simultaneously with the IBIS spectral data, \CaII K images were
obtained through a Halle filter with a nominal FWHM = 0.06 nm,
centered at 0.03 nm from line core.  The spatial scale was set at
0.08\arcsec/pix, and a FOV of approximately 82\arcsec$\times$82\arcsec. Given the overall
low light level with this setup, the exposure time for each image
was 1.2 seconds. Images were obtained every 2 seconds during the same
one hour period, producing a sequence of 1576 frames.

Both sequences of data were subjected to destretching, to remove
residual distortions introduced by the terrestrial atmosphere. For the
narrowband IBIS data, reference images were provided by strictly
simultaneous white light images, while the \CaII K data was
destretched with respect to an averaged image. The \CaII K data were
then interpolated to same spatial scale as the IBIS data, using images
of a reference grid acquired for this purpose. Finally, the \CaII K
filtergram closest in time to the moment of acquisition of the
\CaII 854.2 line core images was selected for each of the 175 spectral
scans.

Fig.~\ref{f_fov} shows the full FOV of the observations, at one time
during the observing sequence. The leftmost panel displays the
intensity image acquired at $-$0.07 nm from the 
\CaII 854.2 line core. As described in Paper I, this position in the line
wing  originates at an average height of $\sim$200 km and the
photospheric reversed granulation pattern 
\citep{2006A&A...450..365J,2007A&A...461.1163C}
 %Janssen & Cauzzi 2006
 % Cheung et al 2007
and magnetic flux concentrations \citep[Paper I;][]{2006A&A...449.1209L} 
% Leenharts 2006
are easily discernible throughout the field. 
The rightmost panel displays instead the
intensity at the nominal \CaII 854.2 line core wavelength formed at a height of
roughly 1200 km. At the center of the
line, fibrils are seen to occupy a large fraction of the FOV, strongly
reminiscent of the structures observed in H$\alpha$. The fibrils can
be seen as both bright and dark elongated structures originating from
even the small magnetic network elements.  For how fibrils appear at
other wavelengths within the line, we refer the reader to Paper I.

The middle panel of Fig.~\ref{f_fov} shows the \CaII K filtergram
closest in time to the displayed IBIS images. 
An essential aspect of this comparison is the presence in the
\CaII K image of features that can be associated with structures present in either
the \CaII 854.2 line wing images (squares), the line core images (ovals), or both (triangles). 
In particular, some 
of the strongest fibrils observed in the \CaII 854.2 core are also visible in the K 
filtergram. 

\section{Data Comparison}\label{s_compare}

The net effect of any instrumental spectral profile is to mix
information from different portions of a spectral line and thus
arising from different layers in the solar atmosphere. This is in
addition to the "smearing" produced from the line formation itself
%%HU Uitenbroek 2006, Sac Peak meeting
\citep{2006ASPC..354..313U}.
Such mixing is particularly relevant for the \CaII H and K
lines because the large passband of the filters generally employed sum
over a significant portion of the entire line profile. In such an
integration, the very low residual intensity of the chromospheric line
center will be swamped by the much higher flux of the line wings
originating in the photosphere. This will strongly dilute the
chromospheric signal when observing on disk. As mentioned in the Introduction, 
in Paper I we put forward
the hypothesis that the scarcity of fibrils observations in H and K
filtergrams is mainly due to this effect.

\begin{figure}
\includegraphics[width=8.4cm,height=6cm]{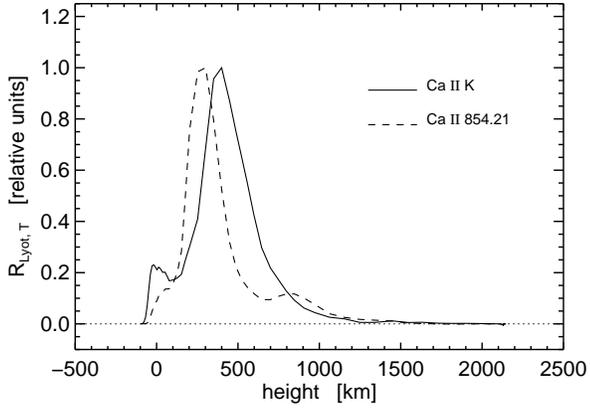}
\caption{Temperature response functions for the \CaII K line (solid)
  and \CaII 854.2 nm line (dashed) after spectral convolution with
  instrumental response (see text for detail). The $x$ axis represents
  heights from the solar surface ($\tau_{500}=1$) in km. Both curves
  have been computed for the FALC model.}
\label{f_response}
\end{figure} 

On the other hand, IBIS data do not suffer this limitation: the
filter passband is comparable to that of a spectrograph, and 
can effectively isolate the chromospheric portions of the line and
resolve fine details in the spectral profiles of \CaII 854.2 (Paper
I and II). Currently, it is not possible to obtain similar measurements in the
H and K lines since no filters with a comparable width
exist in this wavelength range, and it is difficult to obtain
spectroheliograms with similar spatial or temporal
resolution. Therefore, we compare the observations in these two lines
by attempting to replicate the \CaII K filtergram images from the
spectrally resolved \CaII 854.2 data.

\subsection{``Synthetic'' \CaII 854.2  filtergrams}

We introduce a ``synthetic spectral filter'' that, once applied to the
\CaII 854.2 spectra, provides a diagnostic as close as possible to
that of the K data. We adopted two criteria for the construction of
this synthetic filter: 1\,-- that it have the same spectral shape as the
Halle filter, namely a sync function with a large central peak and
secondary transmission orders that introduce signal from the far
wings; 2\,-- that the run with height of the response function to
temperature variations for the two \CaII lines, once convolved with
the respective filter passbands, is as similar as possible. To this
end, we synthetised both spectral lines in the static 1-D FALC model
of the quiet solar atmosphere
\citep[][]{Fontenla+Avrett+Loeser1991}.
% Fontenla+Avrett+Loeser1991.
The best match
for the temperature response function was obtained when using a sync
function of 0.04 nm FWHM, centered at $-$0.025 nm from the core of the
line. Fig.~\ref{f_response} shows the response functions obtained in
this case, for the K line (solid) and the 854.2 line (dashed). Using a
broader passband for the \CaII 854.2 synthetic filter would move the
peak of the temperature response function to much deeper layers, while
positioning it further towards the core of the line would increase the
chromospheric contribution (heights above 800--900 km) well above the
\CaII K case.
Although our choice of filter function is dictated by the hydrostatic
FAL model, this conclusion would not change much if we would have
taken snapshots from chromospheric dynamics simulations as a template,
since the source function of the \CaII 854.2 line is strongly dominated
by the K line
\citep{1989A&A...213..360U},
and the response functions of the two lines would change in similar fashion.

Fig.~\ref{f_sync} shows both spectral lines, together with the
spectral passband of the respective filters (the actual passband
of the Halle filter for \CaII K, and the ``synthetic'' passband for \CaII 854.2). The
top panel further displays the actual IBIS passband at this wavelength, as
well as the spectral positions at which the line was sampled during
the observations. The spectral range plotted for each of the two lines
is proportionate to their wavelength, showing how the position of the
K$_{2V}$ and K$_{2R}$ features in the \CaII K line correspond to the ``knees''
of 854.2 line.
\citet{1989A&A...213..360U} illustrates this scaling and the behavior of the
corresponding line source functions in a quiet-Sun model atmosphere
(his Figure 2).

We then summed the monochromatic images obtained with IBIS after applying
the relative weighting given by the synthetic spectral filter
described above in order to generate ``synthetic filtergrams'' for
the 854.2 line. Only 3\% of the total filter transmission lay outside
of the range of the IBIS spectral scan and was not included in the
construction of the filtergrams.

\subsection{Spatial smearing}\label{spat_smearing}

Both the long exposure times of the \CaII K data (compounded by the
worse seeing at shorter wavelengths), as well some astigmatism
introduced by the Halle filter produce a loss of spatial resolution in
the K images with respect to the IBIS data. The magnitude of the
optical degradation was estimated by comparing the images of the
reference grids taken in the two lines.  The additional seeing
degradation for the K data has been estimated by analyzing the spatial
profile in the \CaII K images of the sharpest bright features visible
in the wings of 854.2, that correspond to isolated magnetic elements.
%\citep[Paper I][]{2006A&A...449.1209L}, 
The composite smearing can be effectively described by an asymmetric 2-D Gaussian with
widths of 1.0 and 1.6 arcseconds  along the two orthogonal axes
of the distribution. The \CaII 854.2 synthetic filtergrams were then convolved
with this PSF to produce spatially smeared images.

\begin{figure*}
\includegraphics[width=5.6cm,height=5.6cm]{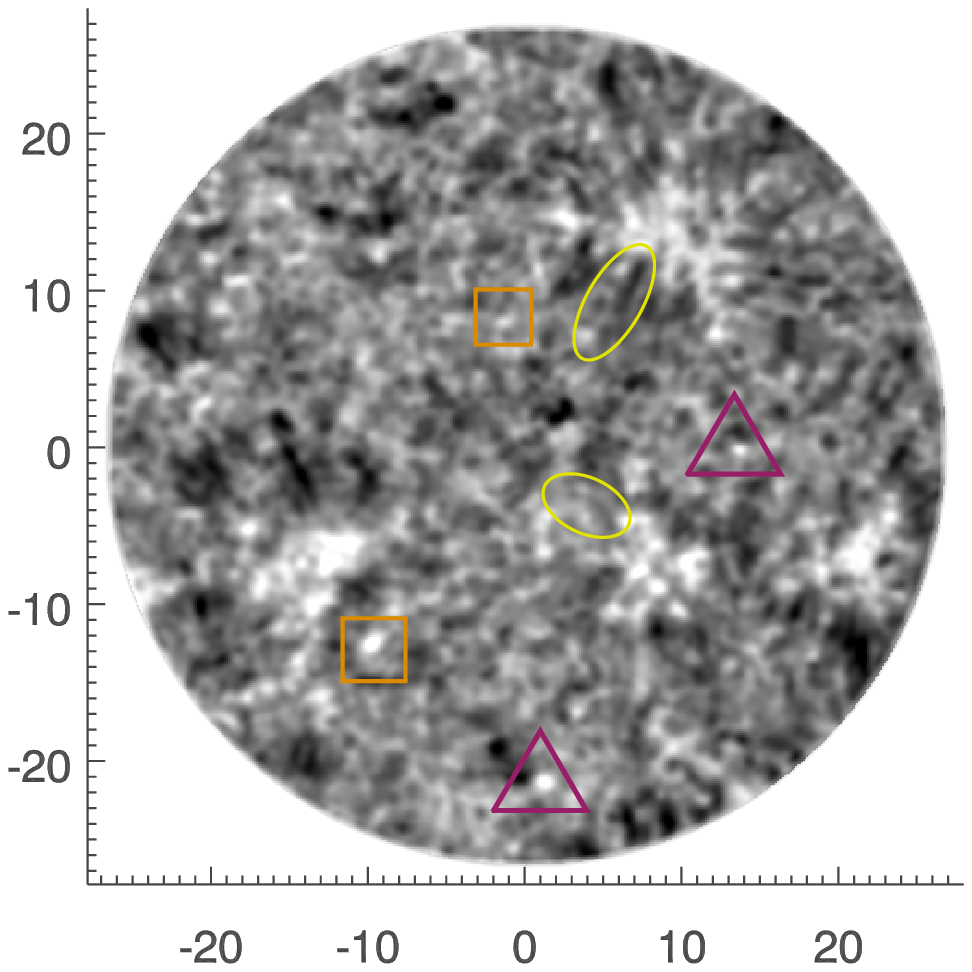}
\hspace{0.4cm}
\includegraphics[width=5.6cm,height=5.6cm]{CaIIK.filtergram.eps}
\caption{Radiation temperature maps derived from the observed K
  filtergram's intensity (left) and synthetic 854.2 filtergram
  (right). The shapes indicate the same features as in Fig. \ref{f_fov}.
  Average values are 4540 and 4600 K, respectively.}
\label{f_trad_maps}
\end{figure*}

\subsection{Radiation temperature}\label{sec:rad_temp}

\begin{figure*}
\includegraphics[width=5.6cm,height=5.6cm]{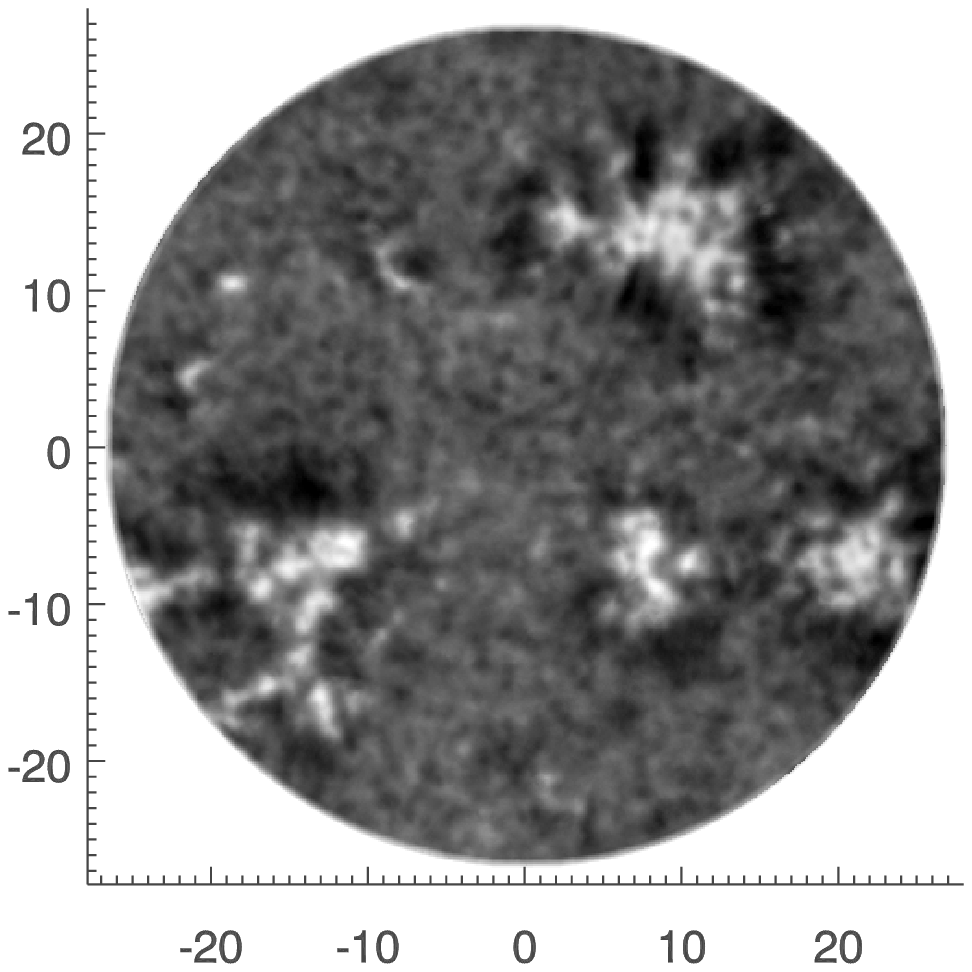}
\includegraphics[width=0.4cm,height=5.7cm]{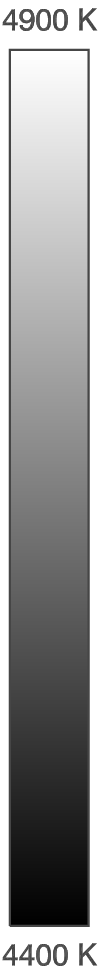}
\includegraphics[width=5.6cm,height=5.6cm]{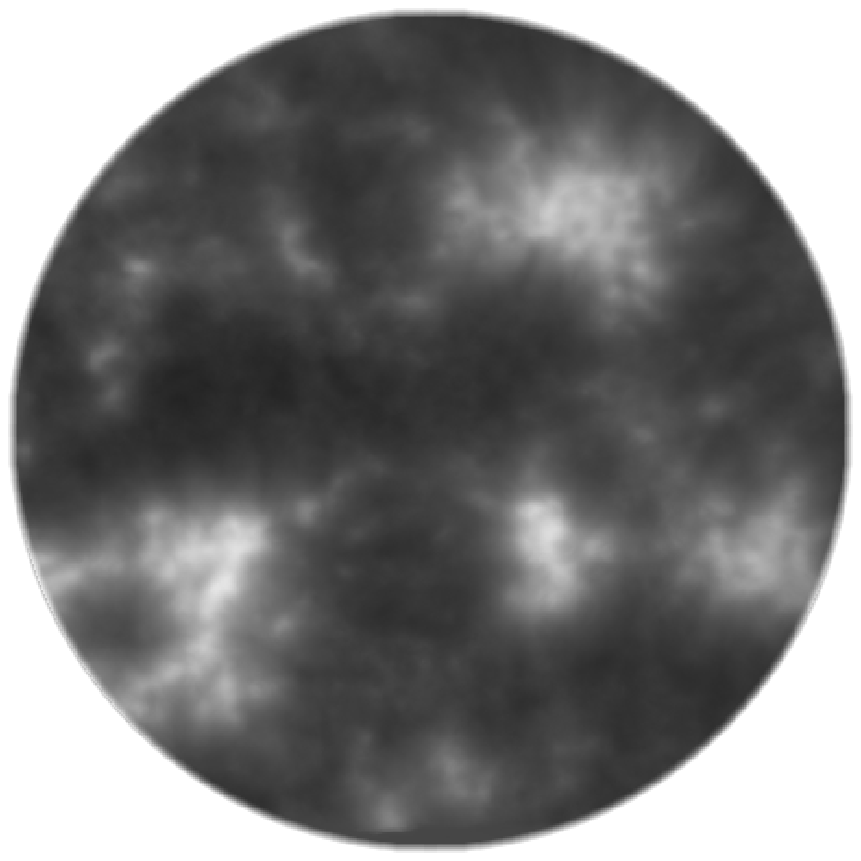}
\includegraphics[width=0.4cm,height=5.7cm]{CaII_K.filt.color_bar.scl1.eps}
\includegraphics[width=5.6cm,height=5.6cm]{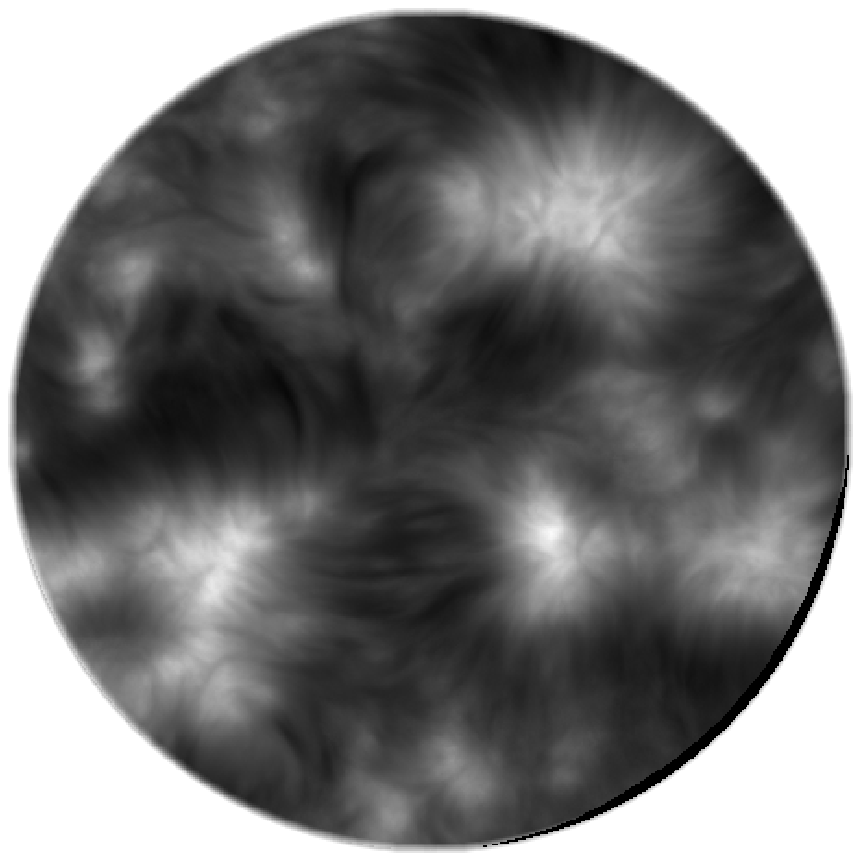}
\includegraphics[width=0.4cm,height=5.7cm]{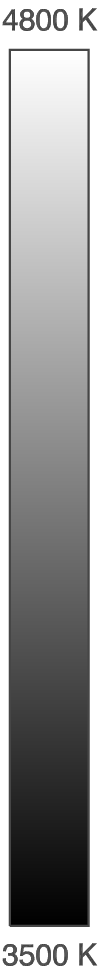}
\caption{Temporally averaged radiation temperature maps summed over
  the 175 realizations obtained during the one hour time series.  
  {\it left:} \CaII 854.2 synthetic filtergrams,
  {\it middle:} \CaII K filtergrams, and
  {\it right:} \CaII 854.2 line minimum intensity. The two filtergrams are both scaled between
  4400 and 4900 K and the line minimum intensity is scaled between 3500 and 4800 K.}
\label{f_trad_aveims}
\end{figure*} 

The intensity variations in the two \CaII lines reflect the different
sensitivity to velocity, density and temperature variations, the
latter mostly due to the different slope of the Planck function at the
two different wavelengths. In order to partially account for this
effect, we translated the intensity maps into radiation temperature
maps, normalizing all the images to the spatio-temporal average
intensity over the entire data sets. The radiation temperature
corresponding to the average intensity has been derived from the line
profiles calculated in the FALC model, after convolution with
the spectral transmission profiles, and corresponds to 4540 and 4600 K
for the \CaII K and 854.2, respectively.
The close match between these two average radiation temperatures
indicates that our choice of synthetic filter for the \CaII 854.2
as derived from the temperature response functions line was indeed proper.
The final result is displayed in Fig.~\ref{f_trad_maps} showing the
radiation temperature map obtained for the observed K filtergram
(left) and the synthetic 854.2 filtergram (right), at the same
temporal step of Fig.~\ref{f_fov}.
The overplotted symbols represent the same features as in Fig.~\ref{f_fov}. 
The \CaII K filtergram now shows smaller amplitude excursions than that
of the synthetic \CaII 854.2 filtergram, because the former is much less sensitive to 
the intensity variations induced by the Doppler effect.
%\citet{1978A&A....70..345C}
%HU Cram 1978
Indeed, the small spectral features in 
the core of the \CaII H and K lines (e.g. K${_2v}$ and K${_2r}$) create multiple 
crosstalk signatures of opposite sign that tend to cancel when summed 
together by the spectral transmission profile, rendering filtergrams in these 
lines highly insensitive to such crosstalk. The more classical absorption profile 
of \CaII 854.2 (at least for an averaged atlas profile) shows a significant 
velocity crosstalk signature even after taking into account the synthetic filter profile 
(also because it happens to be narrower than the \CaII K filter). Taking into account
also the increased amplitude of the Doppler shifts at longer wavelengths, the 
\CaII 854.2 nm synthetic filtergrams are four times more sensitive than the \CaII K 
filtergrams to typical chromospheric velocities.

\section{Results}

\subsection{Temporally Averaged Images}

In Figure \ref{f_trad_aveims}, we display the time-averaged radiation
temperature maps
for the  \CaII 854.2 synthetic filtergrams, the \CaII K filtergrams, and the  \CaII 854.2 line minimum 
intensity. 
This latter is derived from the position of the core as determined for each spectral
profile and removes any effects of
velocities in the conversion to the radiation temperature.

The average T$_{rad}$ map for the line minimum still shows
a wealth of resolved, darker fibrils surrounding the extended bright network,
even in this one hour average. Some of the regions of intermediate intensity
(primarily at the top and bottom of the field) indicate the
fibril-free portions of the internetwork (see Paper II for further details).

The image of the temporally averaged \CaII 854.2
synthetic filtergrams also shows the bright network surrounded by
darker regions of fibrils, though their extent is smaller. The dark streaks surrounding
the network are the result of the strong velocity crosstalk from the upflows
arising from the base of the chromospheric fibrils. 

In the middle image showing the averaged filtergram
again the most obvious features are the areas of bright
network outlining the magnetic field concentrations. In this image,
individual fibrils are not obviously visible, but the network is surrounded by
darker regions that correspond closely to the fibril-dominated
areas in the \CaII 854.2 image. This indicates that the signature of the
fibrils may be present in the \CaII K filtergrams, even if they are not
always resolved as individual structures due to the various effects described
above.

\subsection{Power maps}\label{sec:power_maps}

\begin{figure*}
\center
\vbox{
  \includegraphics[width=17.4cm,height=5.8cm]{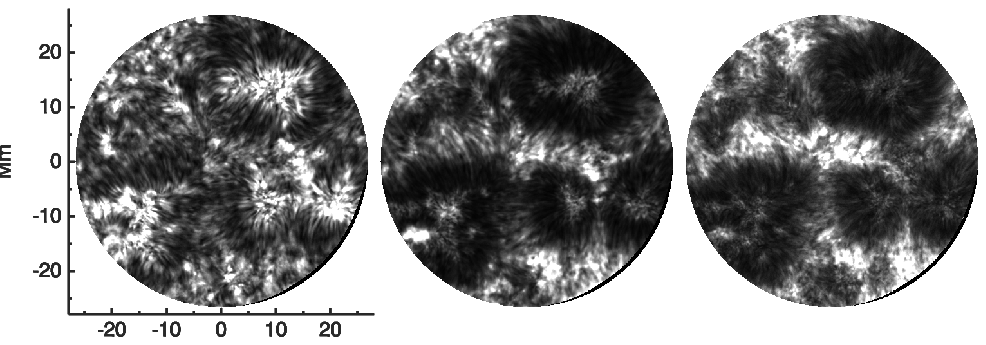}
  \includegraphics[width=17.4cm,height=5.8cm]{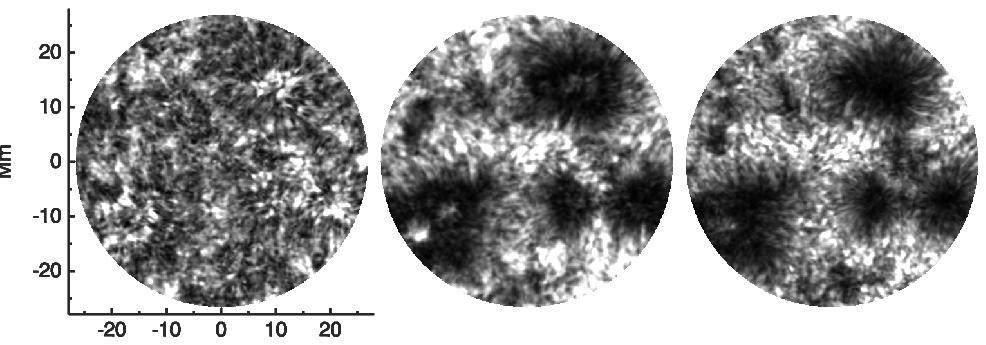}
  \includegraphics[width=17.4cm,height=5.8cm]{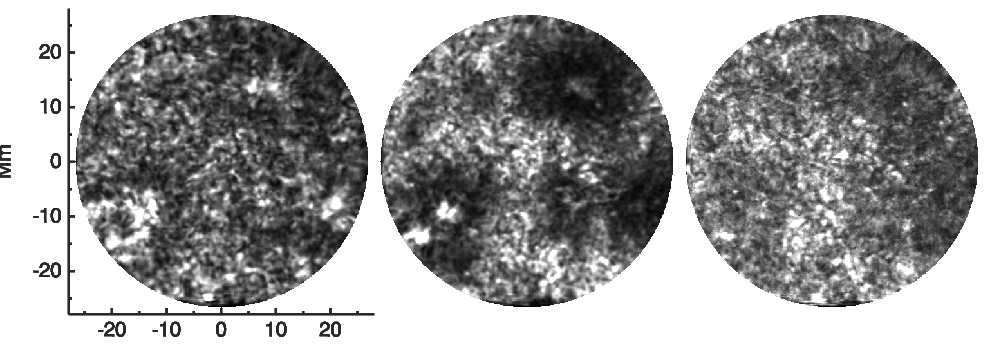}
  \includegraphics[width=17.4cm,height=5.8cm]{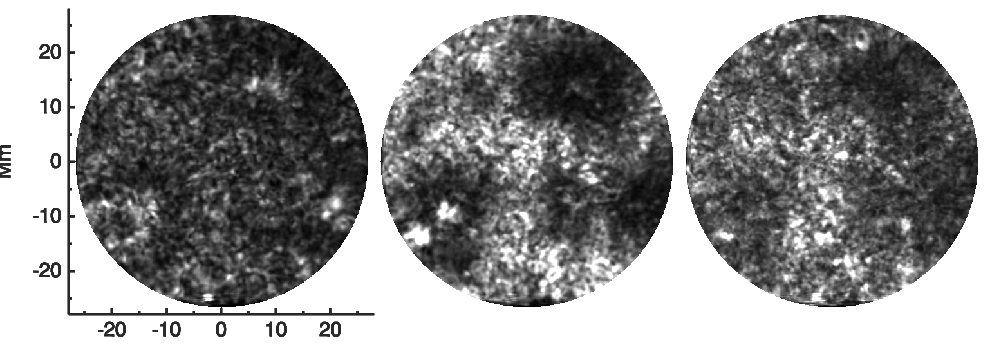}
}
\caption{Power maps from integration of 1-D Fourier power at each pixel over 
three frequency bands arranged in columns ({\it left}) five-minute, 1.8-3.5 mHz;
({\it center}) three-minute, 4-7.5 mHz; and ({\it right}) short period, 7.5-15 mHz. 
The four rows show these maps calculated for \CaII 854.2 line core minimum intensity; 
\CaII 854.2 synthetic filtergrams; 175 \CaII K filtergrams matched to times of synthetic 
filtergrams; 1576 \CaII K filtergrams taken with a strict 2 sec cadence.
}
\label{f_pow_maps}
\end{figure*} 

Since \CaII K images are often used for studies of chromospheric
dynamics using Fourier and other temporal analysis, it is important to
understand to what extent the chromospheric behavior is encoded in
such filtergrams. We therefore compare the Fourier power of the intensity
signals in several different frequency bands.

For each time sequence of radiation temperature maps, we calculate 
the Fourier power of the temperature modulation, separately for
each spatial pixel in the FOV. We normalize by the mean intensity in each pixel in order
to evaluate the modulation power \citep[see ][]{2005A&A...430.1119D}. We
then integrate the power over each of three ranges of frequencies:
1.8--4 mHz; 4--7.5 mHz; and 7.5--15 mHz. 
The calculated power maps are
shown in Figure \ref{f_pow_maps} for the \CaII 854.2 line minimum
intensity (first row), the 854.2 synthetic filtergrams (second row), and the \CaII K
filtergrams (third and fourth row). 

Our ``pure'' chromospheric indicator, the 854.2 line core minimum intensity, shows 
a relative enhancement of the power in the magnetic
network in the five-minute regime and the suppression of the
three-minute oscillations in the fibrils surrounding the network \citep[the ``magnetic shadows'', see][]
{2001ApJ...554..424J}. This is consistent with the behavior found 
by \citet{2007A&A...461L...1V, 2008arXiv0807.4966V} in the
chromospheric velocities for this same dataset.

The same power maps generated using the \CaII 854.2 synthetic filtergrams
and the \CaII K filtergrams show comparable 
behavior in the five-minute and three-minute bands,
with similar spatial distributions of enhanced and suppressed power. 
There are some interesting differences between various power maps, 
such as clearer enhancement of the five-minute power in the network 
in the \CaII K filtergrams as compared to the \CaII 854.2 synthetic
filtergrams. 

We also note that the shadows are more spatially limited and less pronounced in the both 
of the filtergram power maps as as compared to the \CaII 854.2 line minimum power maps.
Using the regions defined in \citet{2008ApJ...683L.207R}, we find that the relative modulation
power in the fibril region is reduced by a factor of six relative to the internetwork power in the 
line core intensities, while the equivalent reductions in the \CaII 854.2 synthetic filtergrams
and the \CaII K filtergrams are only a factor of five and three, respectively.
This shows that the signature of fibrils remains very clear in the power 
maps for the \CaII K filtergrams, even if they are not visible in
the individual intensity images.

The most striking difference in Fig.~\ref{f_pow_maps}
is in the high-frequency regime, where
magnetic shadows are still present in the power maps 
calculated for both the line minimum intensity and 
synthetic filtergrams but are almost nonexistent in the 
\CaII K power map. We test whether this lack of clear
fibril signature might be due to a lower signal-to-noise ratio in the \CaII K filtergrams 
or perhaps caused by the irregular temporal sampling of the \CaII K sequence 
(in order to match them to the closest 854.2 scan). 
We calculate the same power maps using the full 
sequence of 1576 \CaII K images (bottom row in
Fig.~\ref{f_pow_maps}) taken with a regular two-second cadence. 
The shadows are only marginally more visible in this case and the
overall resemblance to the chromospheric power maps 
derived from the \CaII 854.2 line core intensity is still poor. 
It appears that at higher frequencies the \CaII K filtergrams
do not accurately represent the behavior of the chromosphere
itself. 
We note that \citet{1978A&A....70..345C},
%% Cram 1978
using a slit spectrograph,
observed a similar decreased in the high-frequency power in the \CaII K K$_3v$ 
intensity, even though the \CaII 854.2 nm line core 
intensity fluctuations remained significant. This perhaps indicates the lack of high-frequency 
power is due to a line-formation effect, such as the increased opacity of \CaII K leading
to contributions over a larger range of heights, resulting in the cancellation of 
high-frequency oscillations with a mixtures of phases throughout the contribution region.

\subsection{Hinode/SOT \CaII H Filtergrams}\label{sec:sot_filter}

\begin{figure*}
\vbox{
  \includegraphics[width=6.75cm,height=7.75cm]{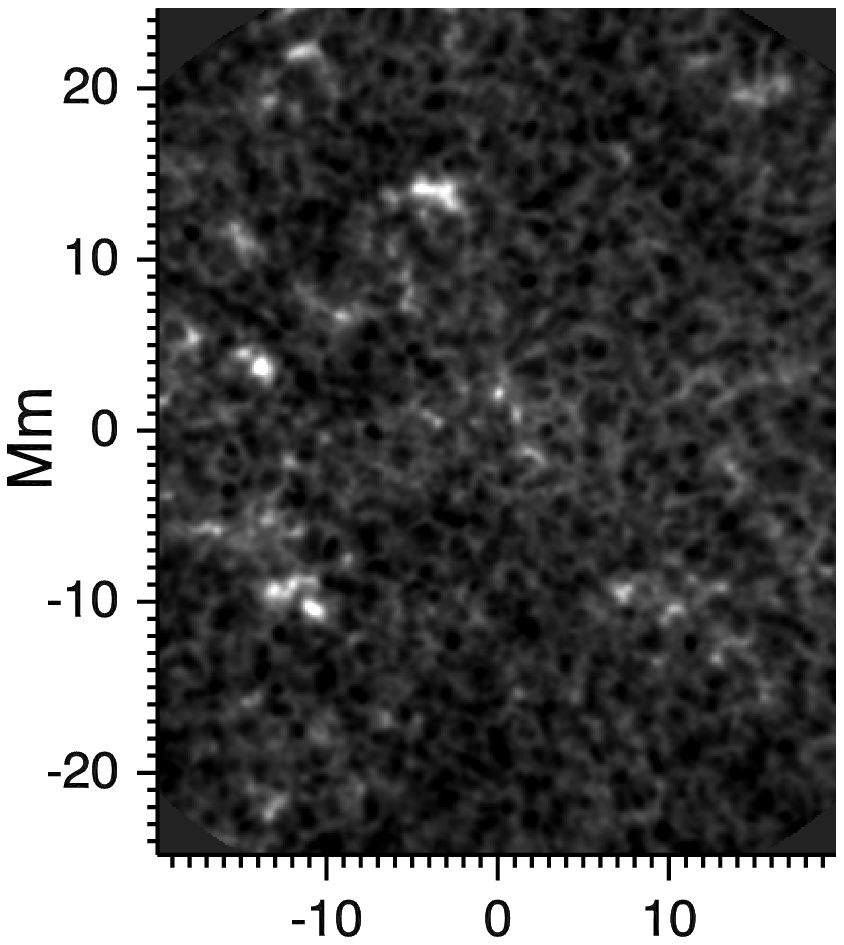}
  \includegraphics[width=5.6cm,height=7.75cm,trim=15mm 0mm 0mm 0mm,clip]{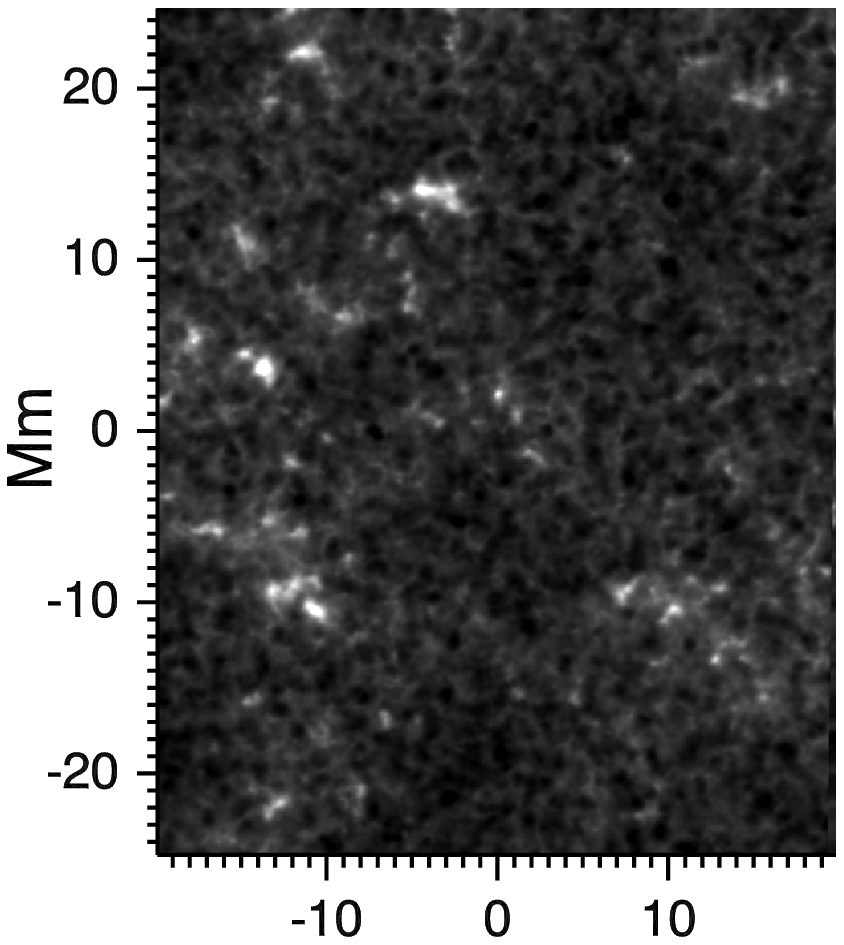}
  \includegraphics[width=5.6cm,height=7.75cm,trim=15mm 0mm 0mm 0mm,clip]{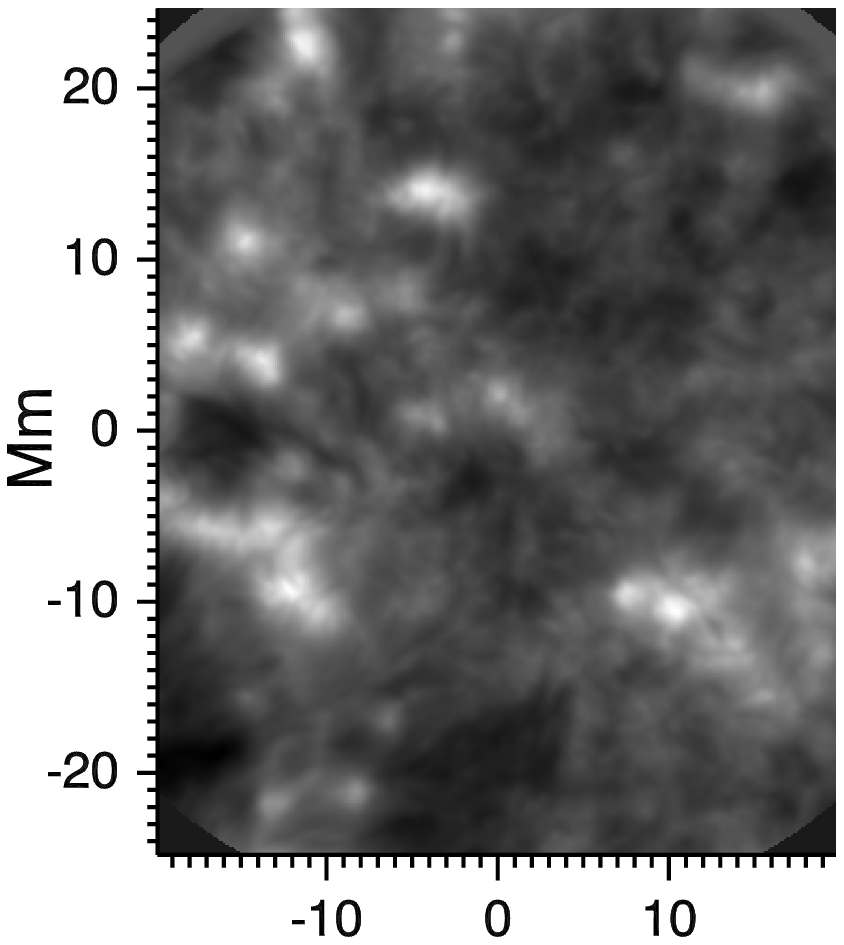}
}
\caption{Observed intensities on 18 April 2008 integrated over a one hour period of ({\it left}) IBIS narrowband image at a position 0.075 nm blueward of 
\CaII 854.2 nm line core (height of formation $\sim$250 km);  ({\it center}) SOT \CaII H filtergram with 0.22 nm FWHM;  and ({\it right}) IBIS narrowband map of line minimum intensity of  \CaII 854.2 (height of formation $\sim$1200 km).}
\label{hinode_int_comp}
\end{figure*} 

\begin{figure*}
\vbox{
  \includegraphics[width=6.75cm,height=7.75cm]{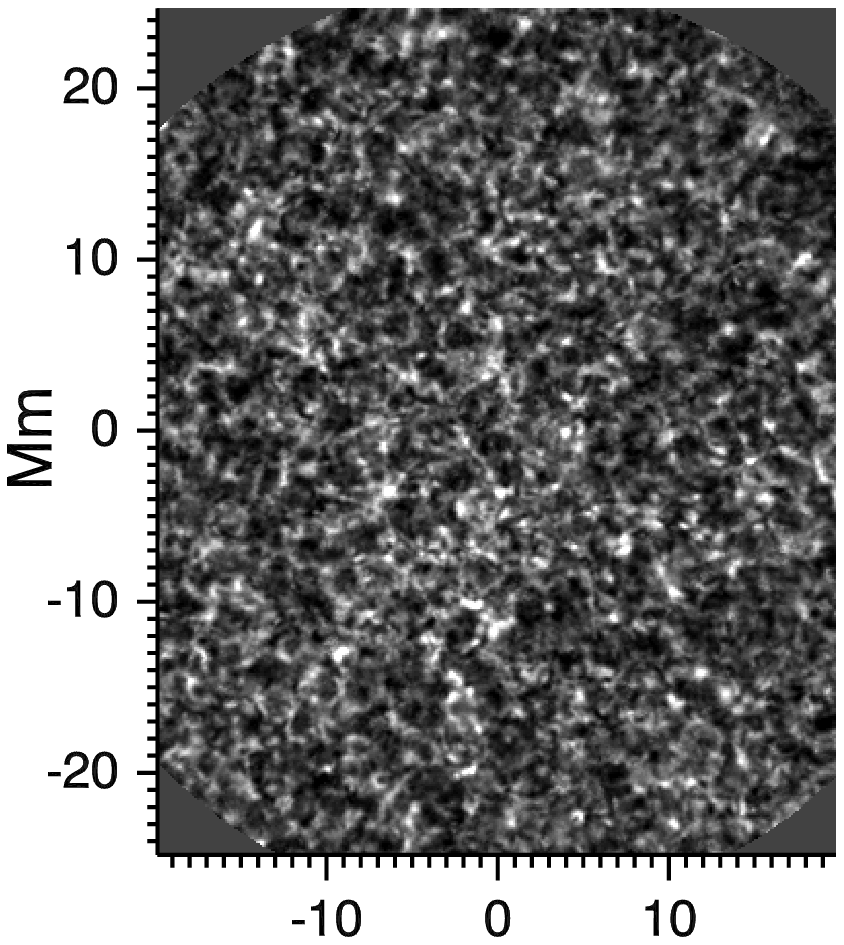}
  \includegraphics[width=5.6cm,height=7.75cm,trim=15mm 0mm 0mm 0mm,clip]{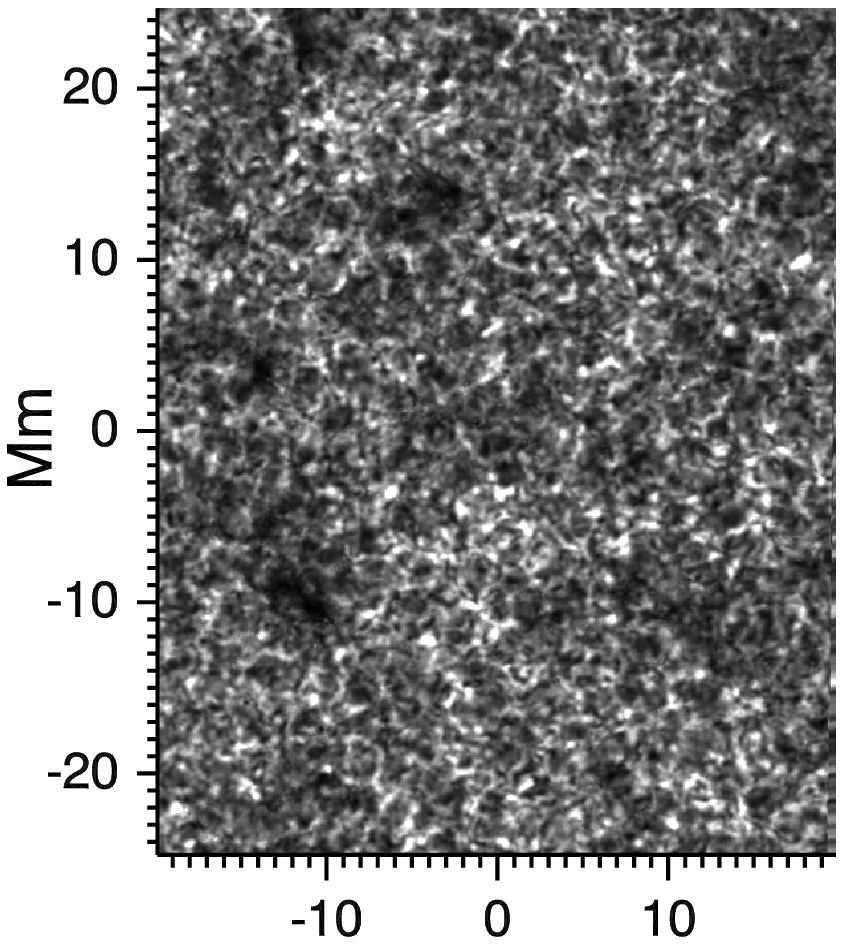}
  \includegraphics[width=5.6cm,height=7.75cm,trim=15mm 0mm 0mm 0mm,clip]{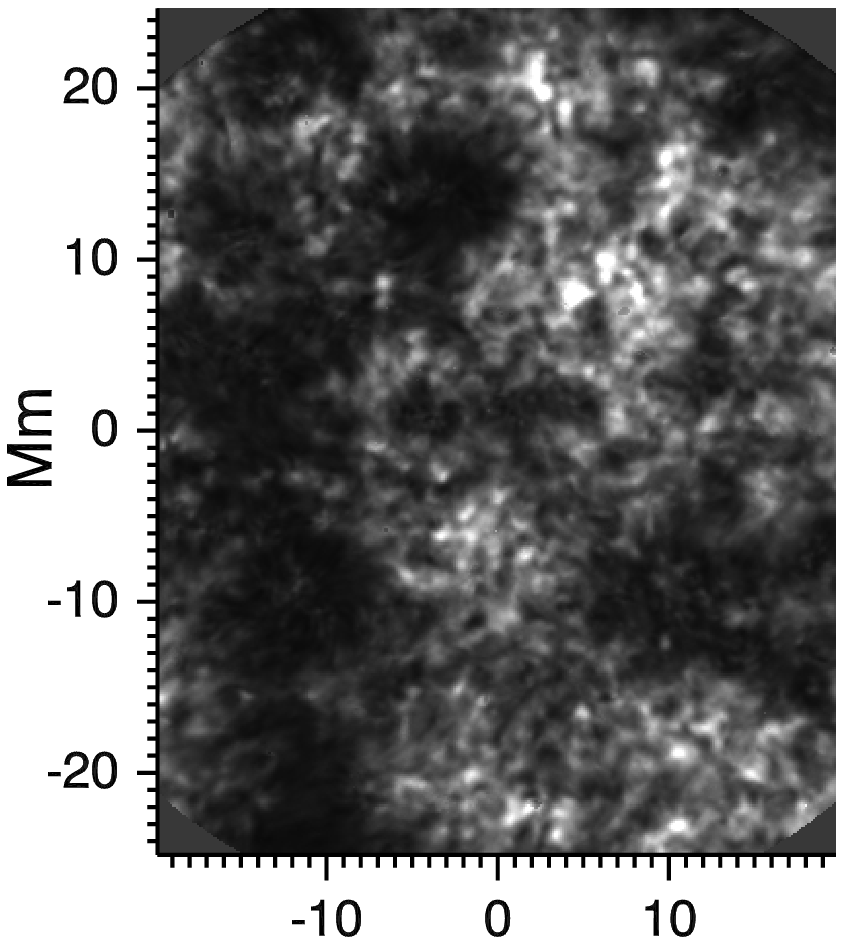}
}
\caption{Power maps from integration of 1-D Fourier power at each pixel over the range 4.5\,-\,7mHz for 
({\it left}) IBIS narrowband images at -0.075 nm from \CaII 854.2 nm line core;  ({\it center}) SOT \CaII H filtergrams;  
and ({\it right}) IBIS line minimum intensity for  \CaII 854.2.}
\label{hinode_pow_comp}
\end{figure*} 

Having identified ways to extract the presence of a chromospheric signal in time series of
filtergrams, we apply these techniques to a sequence of \CaII H filtergrams obtained
with {\it Hinode}/SOT \citep{2007SoPh..243....3K, 2008SoPh..249..167T}. 
% Kosugi 2007 Solar-B Overview
% Tusneta 2008 SOT
We use a different data set, acquired on April 18, 2007 taken near disk center 
in an equatorial coronal hole
We select a one hour interval of simultaneous observations from 16:39\,-\,17:38 UT. 
in which we obtained 500 scans of the \CaII 854.2 line with 
IBIS while the SOT/BFI acquired 176 \CaII H filtergrams.
Because the filter has a FWHM of 0.22 nm, we are not able to apply a similar synthetic filter
to the IBIS data because of the limits of the spectral coverage.
We compute instead the 
average intensity for the \CaII H filtergrams as well as for the \CaII 854.2 line core minimum intensity 
and at a position in the 0.075 nm in blue wing of the line (formed in the middle photosphere 
at an average height of 250 km). 
As seen in Fig.~\ref{hinode_int_comp}, the \CaII 854.2 line wing and \CaII H images are strikingly
similar and the averaged SOT filtergram shows no presence of the chromospheric structures seen
in the line core image. 

As a further test, we also calculated the three-minute power maps in the same way as in 
Sec \ref{sec:power_maps} for the \CaII H filtergrams as well as the \CaII 854.2  line core 
and line wing. Even in this very quiet region the magnetic shadows are again clearly 
visible in the line core intensity power, but there is almost no trace of them in the \CaII H power maps.
The \CaII power map is highly similar, even in the small details, to the photospheric power map 
derived from the wing of the line. Observing on the solar disk, such a broad filter apparently captures 
very little of the chromospheric dynamics.

\section{Conclusions}

In this Paper we examine the causes of the very different 
appearance of on-disk filtergrams obtained with 
the \CaII H or K lines and those that use other chromospheric lines 
such as H$\alpha$ and \CaII 854.2. Given the
widespread use of filtergrams in investigations of the
solar atmosphere, it is important to ask to what extent these kind of data
actually represent the structure and behavior of the chromosphere.

We used a series of repeated imaging spectral scans 
through the \CaII 854.2 line in the quiet chromosphere with IBIS
and compared them with a simultaneous sequence of 
\CaII K filtergrams.
Our main conclusion is that the lack of well defined chromospheric
features in the filtergrams, such as  fibrils and mottles, 
is essentially due to observational limitations. 
We confirmed this by generating synthetic \CaII  ``filtergrams''  by applying 
a suitably chosen simulated filter profile to the spectrally 
resolved 854.2 images. Both 
the observed \CaII K and synthetic \CaII 854.2 intensities were then transformed into radiation
temperature, in order to partially account for the different sensitivity of the Planck function at the 
vastly different wavelengths of the two lines. 

A filter results in the summing of contributions in different points 
throughout a line profile. This is particularly important in the 
\CaII H or K lines given the widths of the filters presently available and 
the broad range of heights over which the line is formed.
Using the nominal shape and position of the Halle filter spectral transmission, 
we derived the temperature response function for the \CaII K line in 
the reference FALC static atmosphere, and showed that it has a broad 
contribution with a peak at around 400 km, 
but equally spanning from the deep photosphere ($\sim$\,100 km) to the middle 
chromosphere ($\sim$\,800 km). 
This effect has long been known and has also been recently
investigated by \citet{2007PASJ...59S.663C}
%Carlsson et al PASJ, Hinode
in a study of high frequency acoustic waves in the quiet chromosphere.
They performed a similar calculation for the \CaII H filter on {\it Hinode}/SOT (with a 0.22
nm FWHM) and found an average response height of 250 km, with a weaker, broad contribution
extending up to 600 km. We confirm this by direct comparison of SOT \CaII H images 
with images from different positions in the \CaII 854.2 line profile and further show that there
is negligible chromospheric signature in the quiet-Sun images taken with the \CaII H filter on SOT.

This summing of contributions results in many chromospheric structures 
being ``swamped'' by the
stronger photospheric signals from the wings of the line.
Dark structures, such as fibrils and filaments, may be most strongly
affected by this mixing, since their contribution becomes increasingly 
negligible compared to the that from the wings of the lines. 
%KR - rewrite
Brighter chromospheric structures instead tend to persist, especially
magnetic structures which show relative brightness enhancement both at chromospheric
heights as well as at the mid-photospheric regions that 
make a significant contribution to the overall filtergram intensity.
This explains many of the typical characteristics of \CaII H and K filtergrams: the bright 
network is clearly defined, but only traces of the strongest chromospheric fibrils are seen.
Away from the magnetic network, the filtergrams contain a mixture of the reversed granulation
arising from the mid-photosphere, enhancements caused by small internetwork magnetic elements, 
and the intermittent brightenings due to acoustic events in the chromosphere. 
There is no way to distinguish whether a 
brightening observed in a \CaII H or K filtergram is of chromospheric or photospheric origin.

Our results are noteworthy for 
at least two reasons: first, we prove that  this effect is
important even for the narrowest available K filters ($\approx$ 0.05 nm FWHM);
and second, we do so by direct comparison with the features 
observed in a subordinate \CaII line directly related to the resonance lines. No chromospheric
model is yet capable of reproducing the spectral characteristics of magnetic 
features in the quiet atmosphere.

Beyond the effect of the filter transmission profile, there is additional degradation of
the chromospheric signal in the \CaII filtergrams due to other observational effects.
The decreased spatial resolution in the \CaII filtergrams compared
to the \CaII 854.2 images can be explained by the significantly longer exposure time for the
filtergrams (due to a combination of dropoff of the Planck function, reduced atmospheric 
transparency, low filter transmission, and decreased detector efficiency) and the worsening of the
atmospheric seeing at the blue wavelengths. 
There are still differences between the observed and synthetic filtergrams, 
primarily the greater visibility of the chromospheric
fibrils in the \CaII 854.2 filtergrams. This is due in large part to the factor of four increase 
in the velocity-to-intensity crosstalk in the \CaII 854.2 synthetic filtergrams due to the narrower
filter profile, the different shapes of the two underlying line profiles, and the increased
magnitude of the Doppler shift at the longer wavelength.

However, even if not immediately visible, the \CaII K filtergrams 
do contain several signatures of the
chromospheric fibrils. These appear
more obvious when taking into account the temporal
evolution. 
Temporally averaged filtergrams show a slightly
decreased intensity surrounding the network magnetic patches 
corresponding to the location of the chromospheric fibrils seen the the 
\CaII 854.2 line core. 

An additional chromospheric signature is the presence of the 
``magnetic shadows'' \citep{2001ApJ...554..424J} in the map of three-minute 
Fourier power derived from the modulation of the radiation temperature in the \CaII K
filtergrams.
These shadows are similar to those seen in the analogous power maps derived from the
\CaII 854.2 line core intensities, but with a more limited spatial extent and an amplitude 
reduced by one half.
Since these shadows correlate with
the presence of the chromospheric fibrils \citep{2007A&A...461L...1V}, this
indicates that the presence of the fibrils can be extracted in the
overall time series of filtergrams.  Such properties can be understood in terms of the different temporal
variability of the various chromospheric components, and in particular, to 
the fact that fibrils present the least intensity (radiation temperature) changes (Paper II).

It is also important to note that the \CaII K filtergrams display little of the structuring seen in
the power maps at higher frequencies using the \CaII 854.2 line. 
In the range 7.5--15 mHz, the map produced
from \CaII 854.2 line core intensity variations shows a continued dependence on the
chromospheric fibril structures, with the magnetic shadows still
prominent. In the maps in the same frequency range from the \CaII K
filtergrams this chromospheric signature is almost completely absent. 
This indicates that on-disk filtergrams in the
\CaII H and K lines, even with relatively narrow filter, may not be
suitable for investigation of the high-frequency behavior of the
chromosphere.

Based on the above discussions, it would appear that the picture given
by the H$\alpha$ and \CaII 854.2 lines, with significant portions of the
chromosphere being dominated by magnetic structuring,
represents the true nature of the chromosphere. As a corollary, on-disk images
that don't show evidence of such structuring, even in the quiet Sun, cannot be said to realistically 
portray the chromospheric conditions.
We expect that images
in \CaII H and K lines taken with higher spatial and spectral
resolution will show a silmilar wealth of fibrillar structures 
\citep{pietarila_espm}. In the absence of such data, and given the observational
advantages of working at longer wavelengths (with the one drawback of
decreased spatial resolution), the \CaII 854.2 nm line represents
an excellent chromospheric indicator.

\acknowledgements{The authors are grateful to the DST observers D. Gilliam, M. Bradford and J. Elrod, whose
patience and skills are greatly appreciated. We thank Rob Rutten for insightful discussions
on the presence and visibility of fibrils in \CaII K.
IBIS was
built by INAF--Osservatorio Astrofisico di Arcetri with contributions 
from the Universit\`a di Firenze and the Universit\`a di
Roma ÒTor VergataÓ. Further support for  
IBIS operation was provided by the Italian MIUR and MAE,
as well as NSO. NSO is operated by the Association of Universities
for Research in Astronomy, Inc. (AURA), under cooperative
agreement with the National Science Foundation. 
This research has made use of NASA's Astrophysics Data System (ADS).

Hinode is a Japanese mission developed and launched by ISAS/JAXA, 
with NAOJ as domestic partner and NASA and STFC (UK) as international 
partners. It is operated by these agencies in co-operation with ESA and NSC (Norway). }

%\begin{thebibliography}
\bibliography{quietchrom,instruments,quietphot,extra}
%\bibliography{k_vs_8542}
%\end{thebibliography}

\end{document}